\documentclass[twocolumn,showpacs,preprintnumbers,amsmath,amssymb]{revtex4}
\usepackage{graphicx}
\usepackage{dcolumn}
\usepackage{bm}

\begin{document}

\title{On Electron Transport in \textrm{ZrB}$_{12}$, \textrm{ZrB}$_{2}$ and \textrm{
MgB}$_{2}$}
\author{V.A. Gasparov}

\author{M.P. Kulakov}
\author{N.S. Sidorov}
\author{I.I. Zver'kova\\}
\affiliation {Institute of Solid State Physics RAS, 142432
Chernogolovka, Moscow District, Russian Federation \\}

\author{V.B. Filipov}
\author{A.B. Lyashenko}
\author{Yu.B. Paderno\\}
\affiliation {Institute for Problems of Material Science NANU, Kiev, Ukraine\\}
\date{today}
\begin{abstract}

We report on measurements of the temperature dependence of resistivity, $\rho(T)$, for single
crystal samples of \textrm{ZrB}$_{12}$, \textrm{ZrB}$_{2}$ and polycrystalline samples of
\textrm{MgB}$_{2}$. It is shown that cluster compound \textrm{ZrB}$_{12}$ behaves like a
simple metal in the normal state, with a typical Bloch - Gr\"uneisen  $\rho(T)$ dependence.
However, the resistive Debye temperature, $T_{R}=300~K$, is three times smaller than $T_{D}$
obtained from specific heat data. We observe the $T^{2}$ term in $\rho(T)$ of these borides,
which could be interpreted as an indication of strong electron-electron interaction.
Although the $\rho (T)$ dependence of \textrm{ZrB}$_{12}$ reveals a sharp
superconductive transition at $T_{c}=6.0~K$, no superconductivity was observed for single
crystal samples of \textrm{ZrB}$_{2}$ down to $1.3~K$.

\end{abstract}
\maketitle

It is known that boron has a tendency to form cluster compounds. In particular there are
octahedral \textrm{B}$_{6}$ clusters in \textrm{MeB}$_{6}$, icosahedral \textrm{B}$_{12}$
clusters in $\beta $-rhombohedral boron, and cubo-octahedral \textrm{B}$_{12}$ clusters in
\textrm{MeB}$_{12}$.  So far, several superconducting cubic hexa - \textrm{MeB}$_{6}$
and dodecaborides - \textrm{MeB}$_{12}$ have been discovered \cite{matt68}
(Me=Sc, Y, Zr, La, Lu, Th). Many other cluster borides (Me= Ce, Pr, Nd, Eu, Gd, Tb, Dy, Ho,
Er, Tm) were found to be ferromagnetic or antiferromagnetic \cite{matt68,pad95}.
Even though the superconductivity in \textrm{ZrB}$_{12}$ was discovered a
long time ago ($T_{c}=6~K$) \cite{matt68}, there has been little effort devoted to the
study of electron transport and basic superconductive properties of dodecaborides.
Only recently, the electron transport of solid solutions
\textrm{Zr}$_{1-x}$\textrm{Sc}$_{x}$\textrm{B}$_{12}$ \cite{hamada93} as well as the
band structure calculations of \textrm{ZrB}$_{12}$ \cite{shein03} has been reported.
Understanding the properties of the cluster borides as well as the superconductivity
mechanism in these compounds is very important.

Recently, we reported superconductivity at $5.5~K$ in the polycrystalline samples of
\textrm{ZrB}$_{2}$  \cite{gasp73}. This was not confirmed in later studies \cite{fisher03}.
It was recently suggested \cite{ivan03} that this observation could be associated with
nonstoichiometry in the zirconium sub-lattice. In this letter we address this problem.
We present the results from measurement of the temperature dependencies of resistivity,
$\rho(T)$, for single crystals of \textrm{ZrB}$_{12}$ and \textrm{ZrB}$_{2}$. Comparative
data from  polycrystalline samples of \textrm{MgB}$_{2}$ are also presented. The
superconducting properties of \textrm{ZrB}$_{12}$ will be published elsewhere.

Under ambient conditions, dodecaboride \textrm{ZrB}$_{12}$ crystallizes in the
\textit{fcc} structure of the \textrm{UB}$_{12}$ type (space group \textit{Fm3m}),
$a=0.7408~nm$ \cite{leithe02}. In this structure, the \textrm{Zr} atoms are located
at interstitial openings in the close-packed \textrm{B}$_{12}$ clusters \cite{hamada93}.
In contrast, \textrm{ZrB}$_{2}$ shows a phase consisting of two-dimensional graphite-like
monolayers of boron atoms with a honeycomb lattice structure, intercalated with \textrm{Zr}
monolayers (with lattice parameters $a=0.30815~nm$ and $c=0.35191~nm$ \cite{gasp73}).

The \textrm{ZrB}$_{2}$ powder was produced by the boron carbide reduction of
\textrm{ZrO}$_{2}$. The \textrm{ZrB}$_{12}$ single crystals were obtained from a mixture
of a certain amount of \textrm{ZrB}$_{2}$ and an excess of boron ($50-95\%$). The resulting
materials were subjected to a crucible-free RF-heated zone-induction melting process in
an argon atmosphere. The obtained single crystal ingots of \textrm{ZrB}$_{12}$ and
\textrm{ZrB}$_{2}$ have a typical diameter of about $5-6~mm$ and a length of $40~mm$.
A metallographic investigation detected that the \textrm{ZrB}$_{2}$ crystal is surrounded by a
polycrystalline rim about $0.5~mm$ thick. The measured specific density of the
\textrm{ZrB}$_{12}$ rod is $3.60~g/cm^{3}$, in good
agreement with the theoretical density. The \textrm{X}-ray diffraction measurements
confirmed that both ingots are single crystal. We found the cell parameters of
\textrm{ZrB}$_{12}$,  $a=0.74072\pm\ 0.00005~nm $, to be very close to the published values
\cite{leithe02}.

Polycrystalline \textrm{MgB}$_{2}$ and \textrm{CaMgB}$_{2}$ samples were sintered
from metallic \textrm{Mg} or a mixture of \textrm{Ca,Mg} powders and boron pellets
using a similar technique as outlined in our earlier work \cite{gasp73}. This technique
is based on the reactive liquid \textrm{Mg,Ca} infiltration of boron. \textrm{X}-ray
diffraction patterns and optical investigation show large grains of single \textrm{MgB}$_{2}$
phase, with much smaller grains of semiconducting \textrm{CaB}$_{6}$ phase visible
in-between. Density of \textrm{MgB}$_{2}$  grains was rather high, $2.4~g/cm^{3}$,
while the samples prepared from \textrm{Mg} infiltration had smaller density of
$2.2~g/cm^{3}$. Only \textrm{MgB}$_{2}$ samples cut from large grains were
studied. These samples will be denoted as \textrm{CaMgB}$_{2}$.

We used a spark erosion method to cut the samples into a parallelepiped with
dimensions of about $0.5\times 0.5\times 8~mm^{3}$. Single crystal samples were oriented along
$<$100$>$  for \textrm{ZrB}$_{12}$, and in hexagonal $[$0001$]$ and basal $[1\bar{1}00]$
directions for \textrm{ZrB}$_{2}$, respectively. The orientation process was performed
using an \textrm{X}-ray Laue camera. The samples were lapped by diamond paste and
subsequently etched: \textrm{ZrB}$_{12}$ in hot nitrogen acid, \textrm{ZrB}$_{2}$
in mixture of \textrm{H}$_{2}$\textrm{O}$_{2}/$\textrm{HNO}$_{3}$/\textrm{HF}, and
\textrm{MgB}$_{2}$ in $2\%$\textrm{HCl} plus water-free ethanol.

A standard four-probe \textit{ac} ($9Hz$) method was used for resistance measurements.
We used \textit{Epotek H20E} silver epoxy for electrical contacts. The samples were mounted
in a a temperature variable liquid helium cryostat. Temperature was measured with platinum
(PT-103) and carbon glass (CGR-1-500) sensors. The critical temperature measured by
RF susceptibility \cite{gasp73} and $\rho(T)$ was found to be  $T_{c0}=5.97~K$ for
\textrm{ZrB}$_{12}$ samples and $39~K$ for \textrm{MgB}$_{2}$ polycrystalline samples,
respectively.

\begin{figure}
\includegraphics[width=8.5cm,height=7.5cm]{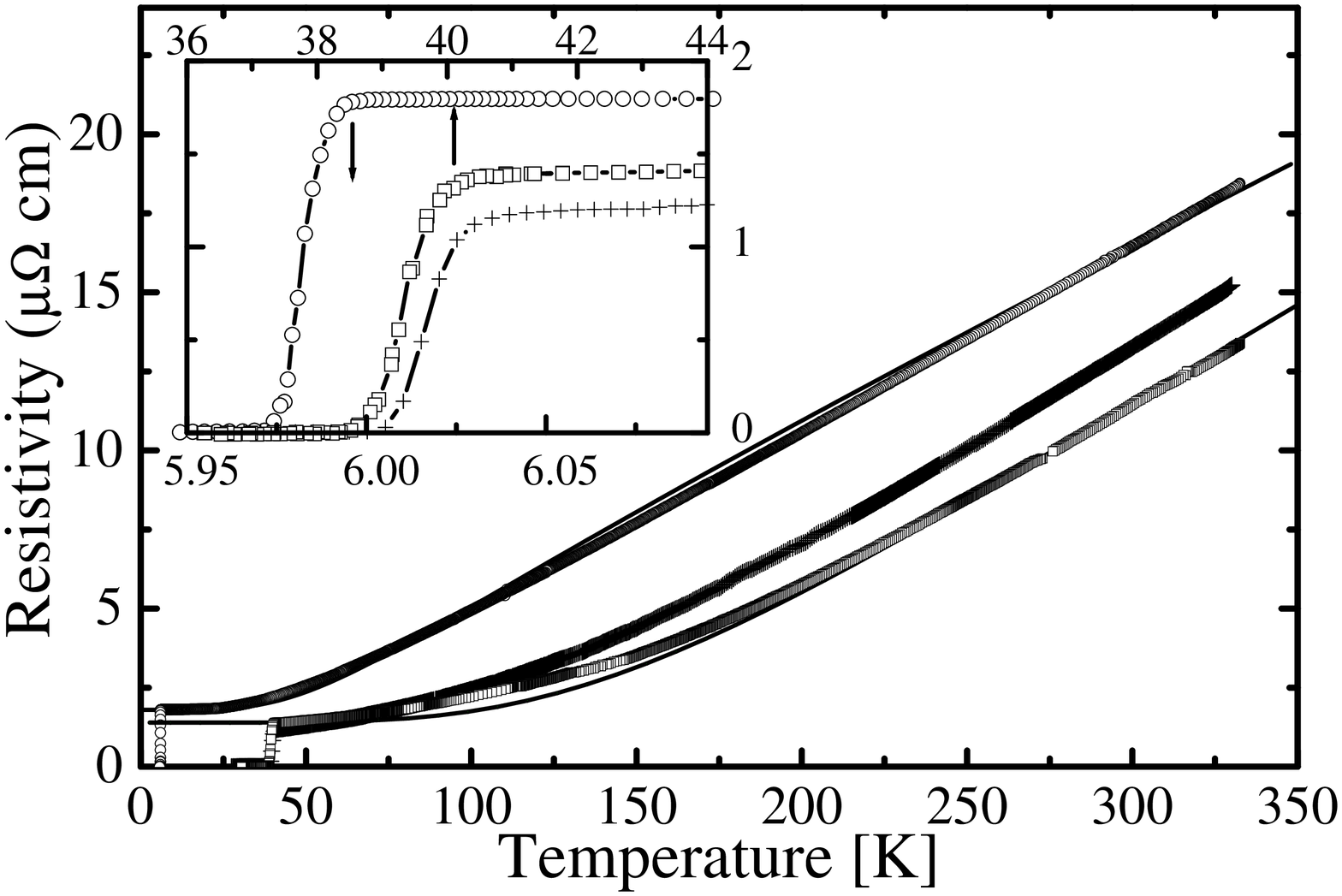}
\caption{Temperature dependence of the resistivity, $\rho(T)$, of \textrm{ZrB}$_{12}$
single crystal (open circles),  \textrm{MgB}$_{2}$ (squares) and \textrm{CaMgB}$_{2}$
(triangles) samples. The solid lines represent BG fits to the experimental data by
Eq.~\ref{eq:one}.}
\label{fig:1}
\end{figure}

\begin{figure}
\includegraphics[width=8.5cm,height=7.5cm]{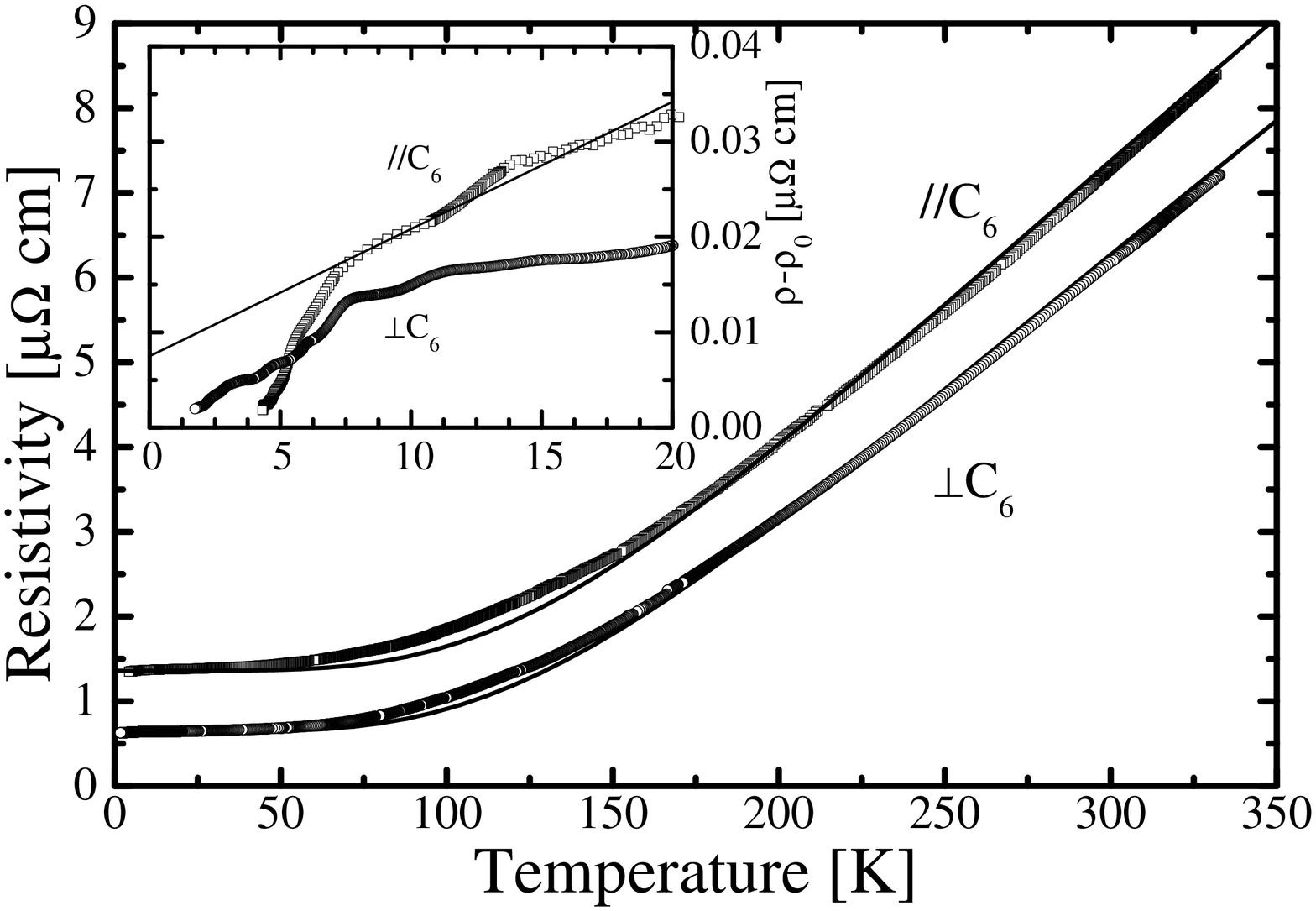}
\caption{Temperature dependence of $\rho (T)$ of \textrm{ZrB}$_{2}$ single crystal samples
in basal plane (circles) and in \textit{c} direction (squares).}
\label{fig:2}
\end{figure}

We display the temperature dependence of the resistivity for \textrm{ZrB}$_{12}$,
\textrm{MgB}$_{2}$ and \textrm{CaMgB}$_{2}$ in Fig.~\ref{fig:1} and that of
\textrm{ZrB}$_{2}$ in Fig.~\ref{fig:2}. To emphasize the variation of
$\rho (T)$ in a superconductive state, we plot these data in  the  inset of  Fig.~\ref{fig:1}.
The samples demonstrate a remarkably narrow superconducting transition with
$\Delta T=0.04~K$ for \textrm{ZrB}$_{12}$ and with $\Delta T=0.7~K$ for both \textrm{MgB}$_{2}$
samples. Such a transition is a characteristic of good quality samples.

As we can see from Fig.~\ref{fig:2}, no superconductivity was observed in \textrm{ZrB}$_{2}$
down to $1.3~K$, while a pronounced slope change in $\rho(T)$ is observed around $7~K$.
One can explain such behavior in the following way. In \textrm{ZrB}$_{2}$ the Fermi level is
located in the pseudo gap.  The presence of \textrm{Zr} defects in \textrm{Zr}$_{0.75}$B$_{2}$
leads to the appearance of a very intense peak in the density of states in the vicinity of the
pseudo gap and subsequent superconductivity \cite{ivan03}. We strongly believe that the
observation of \cite{gasp73} was due to nonstoichiometry of our samples. Superconductivity
in nonstoicheometric samples is very common in other borides: \textrm{MoB}$_{2.5}$,
\textrm{NbB}$_{2.5}$, \textrm{Mo}$_{2}$\textrm{B}, \textrm{W}$_{2}$\textrm{B},
\textrm{BeB}$_{2.75}$ \cite{fisk91,yamam02}.

It is worth noting that \textrm{ZrB}$_{12}$ is mostly boron, and one could speculate that its
resistivity should be rather high. In contrast we observe that the room temperature
resistivity of \textrm{ZrB}$_{12}$ is almost the same as for \textrm{MgB}$_{2}$ and
\textrm{ZrB}$_{2}$ samples. The  $\rho(T)$ is linear above $90~K$ with the slope of $\rho(T)$
more pronounced than in \textrm{MgB}$_{2}$ or \textrm{ZrB}$_{2}$. The residual resistivity
ratio RRR of 9.3 for \textrm{ZrB}$_{12}$ as well as RRR $\approx 10$ for \textrm{MgB}$_{2}$
and \textrm{ZrB}$_{2}$ samples suggests that the samples are  in the clean limit.  One can
predict a nearly isotropic resistivity for \textit{fcc} \textrm{ZrB}$_{12}$, which can
be described by the Bloch-Gr\"uneisen (BG) expression of the electron-phonon
\textit{e-p} scattering rate \cite{ziman60}:

\begin{eqnarray}
\rho(t)-\rho(0)=4\rho _{1}t^{5}\int_{0}^{1/t}\frac{x^{5}e^{x}dx}{(e^{x}-1)^{2}}=
4\rho _{1}t^{5}J_{5}(1/t)
\label{eq:one}
\end{eqnarray}

\begin{figure}
\includegraphics[width=8.5cm,height=7.5cm]{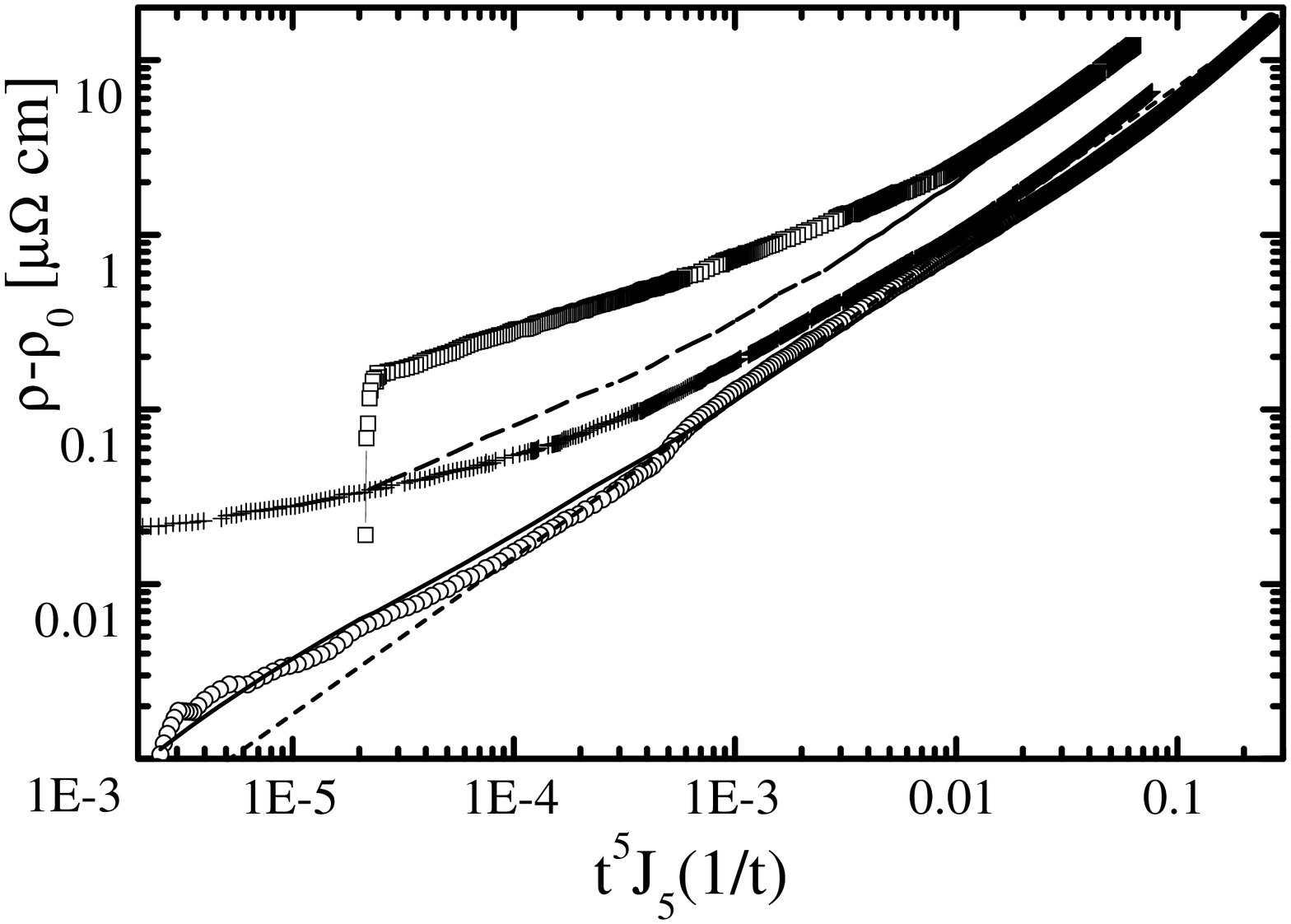}
\caption{The $\rho (T)-\rho (0)$ vs. reduced Debye integral $t^{5}J_{5}(1/t)$ for
\textrm{ZrB}$_{12}$ (open circles), \textrm{ZrB}$_{2}$ in basal plane (crosses) and
\textrm{CaMgB}$_{2}$ (squares). The dashed line is $\rho(T)$ of \textrm{MgB}$_{2}$
calculated in the two band model \cite{mazin02}.}
\label{fig:3}
\end{figure}

Here, $\rho(0)$ is the residual resistivity, $\rho_{1}=d\rho(T)/dT$ is a slope of
$\rho(T)$ at high $T$ ($T>T_{R}$), $t=T/T_{R}$, $T_{R}$ is the resistive Debye
temperature and $J_{5}(1/t)$ is the Debye integral. As we can see from Fig.~\ref{fig:1},
all data for \textrm{ZrB}$_{12}$ fall very close to the theoretical BG function (solid line).
To emphasize the variation of $\rho (T)$ at low $T$ we plot these data as
$\rho(T)-\rho(0)$ versus $t^{5}J_{5}(1/t)$ in Fig.~\ref{fig:3} on a log-log scale.
The BG formula predicts a linear dependence of $log[\rho (T)-\rho (0)]$ versus
$log[t^{5}J_{5}(1/t)]$ with the slope equal to unity. We use $T_{R}$ as a fitting parameter to
achieve agreement at the high temperatures. For comparison, we also present our $\rho(T)$
data of \textrm{ZrB}$_{2}$ and \textrm{MgB}$_{2}$ calculated in a clean case of
the two band model \cite{mazin02}.

It is clear from Fig.~\ref{fig:3} that above $25~K$ the BG model describes the $\rho(T)$
dependence of  \textrm{ZrB}$_{12}$ fairly well. It is remarkable that this description
works well with constant $T_{R}=300~K$. At the same time, $T_{D}$ calculated from
specific heat data \cite{junod04} is three times higher. Furthermore $T_{D}$ increases
from $800~K$ to $1200~K$ as temperature varies from $T_{c}$ up to room temperature.
In order to shed light on this discrepancy, we used a model applied to \textrm{LaB}$_{6}$
of Ref. \cite{mandrus01}. We can treat the boron sub-lattice as a Debye solid with $T_{R}$
and the \textrm{Zr} ions as independent Einstein oscillators with
characteristic temperature $T_{E}$. The effect of the Einstein mode on the resistivity
of a metallic solid is discussed in Ref.\cite{cooper74}:

\begin{eqnarray}
\rho_{E}(T)=\frac{KN\cdot e^{T_{E}/T}}{M\cdot T(e^{T_{E}/T}-1)^{2}}
\label{eq:two}.
\end{eqnarray}

Here $N$ is the number of oscillators per unit volume, $K$ is a constant that depends on
the electron density of the metal, $M$ is the atomic mass. We fit the data by summing
Eq.~(\ref{eq:one}) and  Eq.~(\ref{eq:two}), and living $KN/M$,
$\rho _{1}$, $T_{R}$ as free parameters. Although the model calculations perfectly match
the data (see solid line in Fig.~\ref{fig:3}), the $T_{E}$ we are getting is unreasonably
small ($T_{E}=50~K$), and the difference between $T_{R}$ and specific heat $T_{D}$ becomes
even worse, $T_{R}=270~K$.  We believe that this in-consistency of $T_{R}$ and $T_{D}$ can
be explained by limitation of $T_{R}$ by a cut-off phonon wave vector $q=k_{B}T/ \hbar s$.
The latter is limited by the Fermi surface (FS) diameter $2k_{F}$ \cite{gantm74} rather
than the highest phonon frequency in the phonon spectrum.

According to band structure calculations \cite{shein03}, the FS of \textrm{ZrB}$_{12}$
consists of an open sheet along $\Gamma L$ direction at point $\Gamma$ with $k_{\Gamma
X}=0.47~$\AA$^{-1}$, a quasi spherical sheet at point \textrm{X} ($k_{X\Gamma
}=0.37~$\AA$^{-1}$) and a small sheet at point \textrm{K} ($k_{K\Gamma
}=0.14~$\AA$^{-1}$). We suggest that $T_{R}$ is limited by the small FS sheet.
Unfortunately the experimental FS model and the sound velocity are not yet known.
Therefore we can not corroborate this suggestion by experimental FS.

As we can see from Fig.~\ref{fig:3}, the $\rho (T)$ of \textrm{ZrB}$_{2}$ and
\textrm{MgB}$_{2}$ samples deviates from the BG model even more dramatically.
Putti \textit{et al.} \cite{putti02} modified the BG equation introducing variable
power \textit{n} for the $t^{n}J_{n}(1/t)$ term in Eq.~(\ref{eq:one}). The best
fit to the data was obtained with \textit{n} = 3 which in fact ignores a small
angle \textit{e-p} scattering. Recently Sologubenko \textit{et al.}
\cite{solog02} reported a cubic $T$ dependence in the \textit{a,b} plane resistivity below
$130~K$ in the single crystals of \textrm{MgB}$_{2}$. This was attributed to the
interband \textit{e-p} scattering in transition metals.

However, we believe there are strong objections to this modified BG model: (i) a cubic
$\rho (T)$ dependence is a theoretical model for large angle \textit{e-p} scattering and
no evidence of it was  observed  in transition and non-transition metals;
(ii)  the numerous studies of the $\rho (T)$ dependence in transition metals have
been found to be consistent with a sum of electron-electron \textit{e-e}, $T^{2}$,
and \textit{e-p}, $T^{5}$, contributions to the low $T$ resistivity, which may easily be
confused with a $T^{3}$ law \cite{ziman60,volk73,gasp93}; (iii) the
interband $\sigma -\pi$ \textit{e-p} scattering plays no role in normal transport in
the two band model for \textrm{MgB}$_{2}$ \cite{mazin02}.

\begin{figure}
\includegraphics [width=8.5cm, height=12cm] {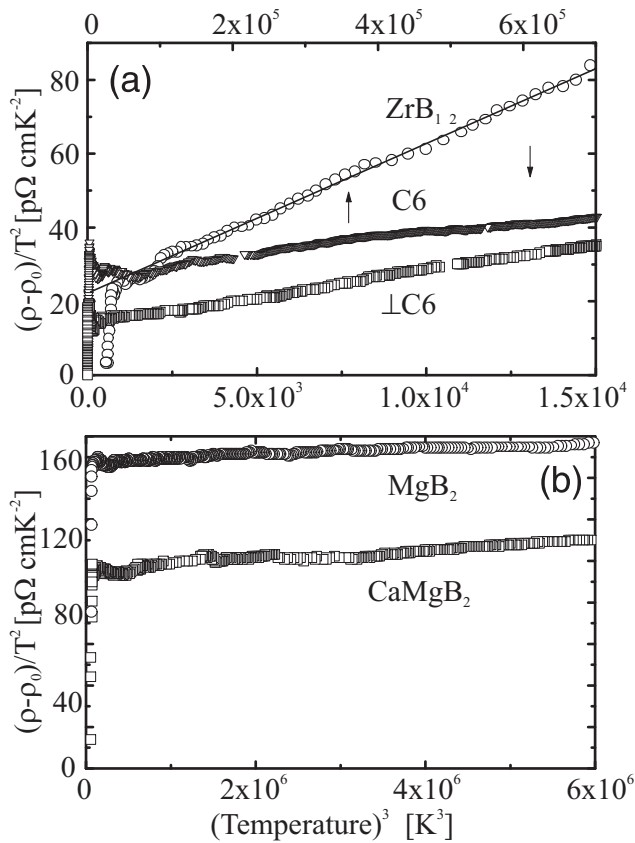}
\caption{Low temperature behavior of $[\rho(T)-\rho(0)]/T^{2}$ versus $T^{3}$ for:
(a) \textrm{ZrB}$_{12}$ (circles), \textrm{ZrB}$_{2}$ in basal plane (squares),
\textrm{ZrB}$_{2}$  along \textit{c} (triangles) and (b) \textrm{MgB}$_{2}$ (circles)
and \textrm{CaMgB}$_{2}$ (squares) samples.}
\label{fig:4}
\end{figure}

In order to solve these problems, we added \textit{e-e} scattering $T^{2}$ term in
Eq.~(\ref{eq:one}) \cite{volk73,gasp93} as a possible scenario.  Indeed, keeping in
mind that the BG term is proportional to $T^{5}$ at $T<0.1T_{R}$, $\rho (T)$ dependence
may be presented in a  simple way \cite{volk73,gasp93}: $[\rho(T)-\rho(0)]/T^{2}=
\alpha +\beta T^{3}$. Here $\alpha$ and $\beta =497.6\rho_{1}/T_{R}^{5}$  are parameters
of \textit{e-e} and \textit{e-p} scattering terms, respectively. Such a plot should
yield a straight line with slope of $\beta$ and its intercept with y-axis ($T=0$)
should equal to $\alpha$. Further, to be consistent with BG law, the $\beta$ parameter
should lead to the same $T_{D}$ as obtained from high $T$ log-log fit in Fig.~\ref{fig:3},
and both coefficients must be independent of $\rho(0)$. We determined $\rho(0)$ from
the intercept of linear $\rho (T)$ vs $T^{2}$ dependence with the $T=0$ axis and plotted
the [$\rho(T)-\rho(0)]/T^{2}$ vs. $T^{3}$ in Fig.~\ref{fig:4}. It is evident that
the measured resistivity approaches a quadratic law at $T<25~K$ in \textrm{ZrB}$_{12}$,
at $T<100~K$ in \textrm{ZrB}$_{2}$, and at $T<150~K$ in both \textrm{MgB}$_{2}$ samples.

The regime of applicability of two term fit is limited to temperatures below
$0.1T_{R}$. At larger $T$ the \textit{e-p} term increases more slowly than $T^{5}$ law
and this is why the data are not consistent any more with the two terms equation.
From the intercept with $T=0$ axis, we find very similar values of $\alpha$ for
\textrm{ZrB}$_{12}$ and \textrm{ZrB}$_{2}$ samples in the basal plane
($\alpha =22~p\Omega cmK^{-2}$ and $15~p\Omega cmK^{-2}$, respectively) while $\alpha $
is about five times larger for \textrm{CaMgB}$_{2}$ sample, $95~{p\Omega cmK^{-2}}$.
The slopes of $\beta $ give $\rho_{1}$ and $T_{R}$ values largely consistent with
high temperature log-log fits for the \textrm{ZrB}$_{12}$ and \textrm{ZrB}$_{2}$ samples.

However, low $T$ results for $\beta$ and $\rho _{1}$ are far from consistent with
high $T$ data for both \textrm{MgB}$_{2}$ and \textrm{CaMgB}$_{2}$ samples. Nevertheless,
the magnitude of $T_{R}=900~K$ for \textrm{MgB}$_{2}$ extracted from log-log fit above
$150~K$, is in excellent agreement with $T_{D}=920~K$ obtained from low-temperature
specific heat measurements \cite{junod02}, and is considerably lower then the reported
data based on $T^{3}$ dependence of $\rho (T)$ ($T_{R}=1050-1226~K$, where $T^{2}$ term
was ignored.  \cite{fisher03,putti02,solog02}).  A similar fit for theoretical
curve is even more consistent with
$T_{R}=900~K$, however we have to mention that violation of Matthiessen's rule in
\textrm{MgB}$_{2}$ may mask the intrinsic  $\rho(T)$ dependence \cite{mazin02}.

In general, there are many scattering processes responsible for the $T^{2}$  term in
$\rho(T)$ of metals: (i) size, surface, dislocation and impurity scattering induced
deviations from Matthiessen's rule (see references in \cite{maas90}); (ii) \textit{e-p}
scattering for small cylindrical FS sheets relative to the phonon wave vector \cite{gantm74};
(iii) inelastic electron impurity scattering (\textit{e-i}) \cite{kagan66}; (iv) the
quantum interference between \textit{e-i} and \textit{e-p} scattering \cite{reizer92};
(v) \textit{e-e} scattering \cite{volk73,gasp93}.

We can estimate some of these effects. We use Drude law to obtain the residual
electron mean free path $l=4\pi v_{F}/\rho \omega _{p}^{2}$. Using a Fermi velocity of
$v_{\sigma}=3.2\cdot 10^{7}~cm/s$ and a plasma frequency
$\omega _{p}^{\sigma }=5.16\cdot 10^{15}~s^{-1}$ for \textrm{MgB}$_{2}$
$\sigma$-band \cite{mazin02}, we obtain $l\approx 100~nm$. This implies that size effects
are negligible for both \textrm{MgB}$_{2}$ samples and \textrm{Zr} borides.
In agreement with \textrm{ZrB}$_{2}$  data (see Fig.~\ref{fig:4}) the $\alpha$ is
proportional to $\rho(0)$ for inelastic \textit{e-i} scattering \cite{kagan66,reizer92}.
However, this term is 1.5 times lower for \textrm{CaMgB}$_{2}$ relative to
\textrm{MgB}$_{2}$, which has the same $\rho(0)$.

We can try to estimate contribution from the small FS sheets to $\alpha$. The $T^{2}$
term was observed in $\rho(T)$ and electron scattering rates of \textrm{Bi} and
\textrm{Sb}, which was attributed to a missing of one $q$ component for \textit{e-p}
scattering on small cylindrical FS sheets \cite{gantm74}. The FS of \textrm{MgB}$_{2}$
is composed of two warped open cylinders running along the \textit{c} axis, which
arise from $\sigma $ boron orbitals \cite{mazin02,car03}. The FS of \textrm{ZrB}$_{2}$
consist of nearly ellipsoidal surfaces joined together at the corners
\cite{tanaka78,rosner02}, which may also be responsible for the $T^{2}$ term in $\rho
(T)$. We can use the sound velocity $s=1.1\cdot 10^{6}~cm/s$ and $8\cdot 10^{5}~cm/s$
for \textrm{MgB}$_{2}$ and \textrm{ZrB}$_{2}$, respectively \cite{shukla03,aizawa01},
to estimate the lowest temperature, $T_{min}=\hbar k_{F}s/k_{B}$, when the phonon wave
vector $q$ matches a neck of smaller $\sigma $ tube in \textrm{MgB}$_{2}$
($k_{\sigma}=0.129~$\AA$^{-1}$  \cite{car03}) or a diameter of the ellipsoidal sheets
in \textrm{ZrB}$_{2}$ ($k_{F}=0.095~$\AA$^{-1}$) \cite{tanaka78}). We obtain
$T_{min}=95~K$ and $60~K$, respectively. Thus we conclude that $q<k_{F}$ at $T<100~K$
in both diborides, which implies that the contribution of the 2D FS sheets to $\alpha
$ is negligible.

In general only umklapp \textit{e-e} scattering contributes to $\rho(T)$, whereas the normal
collisions are significant in compensated metals and in thermal resistivity \cite{gasp93}.
Borides have rather high $T_{D}$ which depresses the \textit{e-p} scattering, so that the
\textit{e-e} SR term is easier to observe. Notice however that the $\alpha$ value for
\textrm{MgB}$_{2}$ is five times larger than corresponding values in \textrm{ZrB}$_{12}$ and
\textrm{ZrB}$_{2}$. The latter values are in turn five times larger than in transition
metals ($\alpha _{Mo}=2.5~p\Omega cm/K^{2}$ and $\alpha _{W}=1.5-4~p\Omega cm/K^{2}$
\cite{volk73,gasp93}). Therefore, additional experiments must be performed for more pure
samples before final conclusion  about the origin of  the $T^{2}$ term in borides can be drawn.

In conclusion, we present a study of the $\rho(T)$ of single crystals of \textrm{ZrB}$_{12}$,
\textrm{ZrB}$_{2}$ and polycrystalline samples of \textrm{MgB}$_{2}$. Large differences
between resistive and specific heat Debye temperatures have been observed for
\textrm{ZrB}$_{12}$. The results provide evidence of a $T^{2}$ term for all these borides
at low $T$, whose origin is not yet understood.

\begin{acknowledgments}
Very useful discussions with V.F. Gantmakher, A. Junod, I. Shein, R. Huguenin,  and
help in paper preparation of L.V. Gasparov are gratefully acknowledged. This work was
supported by the Russian Scientific Programs: Superconductivity of Mesoscopic and Highly
Correlated Systems (Volna 4G); Synthesis of Fullerens and Other Atomic Clusters (No.541-028);
Surface Atomic Structures (No.4.10.99), Russian Ministry of Industry, Science and Technology
(MSh-2169.2003.2), RFBR (No.02-02-16874-a) and by the INTAS (No.01-0617).
\end{acknowledgments}

\end{document}